# Brain Emotional Learning-Based Prediction Model

For Long-Term Chaotic Prediction Applications

M. Parsapoor

*Abstract—* **This study suggests a new prediction model for chaotic time series inspired by the brain emotional learning of mammals. We describe the structure and function of this model, which is referred to as BELPM (Brain Emotional Learning-Based Prediction Model). Structurally, the model mimics the connection between the regions of the limbic system, and functionally it uses weighted k nearest neighbors to imitate the roles of those regions. The learning algorithm of BELPM is defined using steepest descent (SD) and the least square estimator (LSE). Two benchmark chaotic time series, Lorenz and Henon, have been used to evaluate the performance of BELPM. The obtained results have been compared with those of other prediction methods. The results show that BELPM has the capability to achieve a reasonable accuracy for long-term prediction of chaotic time series, using a limited amount of training data and a reasonably low computational time.**

## I. INTRODUCTION

Prediction models have applications in many different areas of science and technology: business, economics, healthcare, and welfare services. Well-known data-driven methodologies such as neural networks and neuro-fuzzy models have shown reasonable accuracy in the nonlinear prediction of chaotic time series [1]. According to the Vapnik-Chervonenkis (VC) theory [2], a sufficient size of training samples for achieving arbitrary prediction accuracy is proportional to the number of the models' learning parameters. Thus, a data-driven model with high model complexity and high number of learning parameters requires a large number of training samples to achieve high accuracy in a prediction application. The prediction accuracy of predicting chaotic time series depends on the characteristics of the applied models (i.e., the model complexity and the number of learning parameters) as well as the features of the prediction applications (i.e., the number of training samples, chaos degree, embedding dimension, and the horizon of prediction) [1]–[5].

This study suggests a prediction model that could be feasible for long-term prediction of chaotic systems (chaotic time series) with a limited amount of training data samples; i.e., a prediction model with a small number of learning parameters. The suggested model is inspired by the brain emotional processing and is therefore named BELPM that stands for Brain Emotional Learning-Based Prediction Model. Its simple architecture has been inspired by the brain emotional system, emphasizing the interaction between those parts that have significant roles in emotional learning. The model mimics the emotional learning by merging weighted k-Nearest Neighbor (Wk-NN) method and adaptive neural networks. The learning algorithm of BELPM is based on the steepest descent (SD) and the least square estimator (LSE). The model aims at continuing the recent studies that have suggested computational models of emotional processing for control and prediction applications [6]-[16].

The rest of the paper is organized as follows: Section II gives an introduction of emotional processing in the brain, reviews the anatomical aspects of emotional learning, and gives an overview of related work in the area of computational models of brain emotional learning. Section III describes the BELPM's architecture and illustrates its learning algorithm. In Section IV, two benchmark chaotic time series, Lorenz and Henon, are used to evaluate the performance of BELPM and the results are compared with the obtained results from other nonlinear learning methods (e.g. Adaptive-Network-Based Fuzzy Inference System (ANFIS) [3], MultiLayer Perceptron (MLP) Network [1], [2], Radial Bias Function (RBF) Networks [1], [2], and Local Linear Neuro-Fuzzy (LLNF) Models [1]). Finally, conclusions about the BELPM model and further improvements to the model are stated in Section V.

## II. BACKGROUND

One of the most challenging topics in machine learning research area is development of high generalization algorithms for accurate prediction of chaotic systems. Recently, bio-inspired models, in particular, emotion-based learning models [6], [8], [10], [12]-[14], have shown acceptable generalization capability in modeling and predicting the chaotic behavior of dynamic systems. In fact, this capability is obtained in emotion-based learning models by integrating machine learning algorithms with the computational model of emotional learning. In the following, we explain how emotional learning can be modeled as a computer-based tool and how it can be integrated with learning algorithms. The difference between BELPM and several well-known data-driven methods will also be indicated.

*A. Related Works in Modeling Emotional Learning*

Since 1988, emotion and emotional processing have been active research topics for neuroscientists and psychologists. A lot of efforts have been made to analyze emotional behavior and describe emotion on the basis of different hypotheses, e.g., psychological, neurobiological, philosophy, and learning hypothesis. These hypotheses that have contributed to the present computer-based models of emotional processing [18], have imitated certain aspects of emotional learning, and can be classified on the basis of their fundamental theories and applications. For example, a computer-based model that is based on the central theory [18] (i.e., which explains how a primary evaluation of emotional stimuli forms emotional experiences) is called a computational model of emotional learning and imitates the associative learning aspect of emotional processing [18] that is based on fear conditioning [19], [20].

Emotional processing has also been described using different anatomical structures: "MacLean's limbic system, Cannon's structure, and Papez circuit" [21]. The first anatomical representation is based on studies on cognitive neuroscience [21], [22], and has been developed on the basis of Cannon's structure and Papez circuit emphasizing the role of the limbic system (i.e., a group of the brain regions) for emotional processing. The Cannon's structure suggested that the hypothalamus of the brain plays the most significant role in emotional learning, while Papez circuit [21] emphasized the role of the cingulate cortex in emotional processing. In this study, we have focused on the limbic system, which is the basis of our suggested model.

*1) Anatomical Structure of Emotional Learning*

The limbic system is a group of brain regions, which includes the hippocampus, amygdala, thalamus, and sensory cortex [22]-[24]. The roles of the main regions of the limbic system with regard to emotional learning can be summarized as follows:

*a) Thalamus* receives emotional stimuli and is responsible for the provision of high-level information, i.e., determining the effective values of stimuli [22]-[28]. It then passes the generated signals to the amygdala and sensory cortex [28]. The thalamus includes different parts that process the emotional stimuli separately [27].

*b) Sensory cortex* is a part of the sensory area of the brain and is responsible for analysis and processing of the received signals. The sensory cortex distributes its output signals between the amygdala and orbitofrontal region [18]-[21], [27].

*c) Amygdala* is the central part of the limbic system of mammals and has a principal role in emotional learning [18]-[26]. The amygdala consists of several parts with different functional roles (see Fig. 1), and it connects through them to other regions of the brain (e.g., the insular cortex, orbital cortex, and frontal lobe). It has connections to the thalamus, orbitofrontal cortex, and hypothalamus [25], [26]. During emotional learning, the amygdala participates in reacting to emotional stimuli, storing emotional responses [29], evaluating positive and negative reinforcement [30], learning the association between unconditioned and conditioned stimuli [19], [20], [31], predicting the association between stimuli and future reinforcement [31], and forming an association between neutral stimuli and emotionally charged stimuli [30].

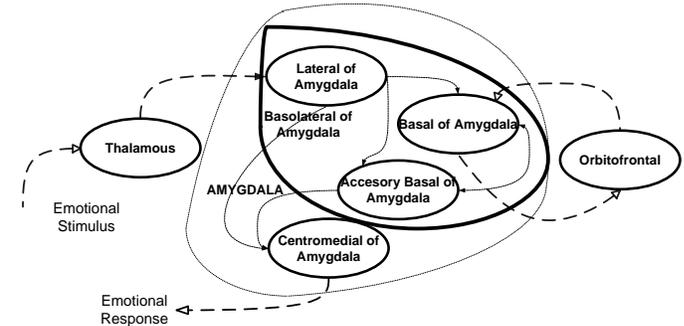

Fig.1.The parts of amygdala and their pathways. The diagram shows the pattern of fear conditioning, which needs to be clarified [17].

The two main parts of the amygdala are the basolateral part (the largest portion of the amygdala) and the centeromedial part. The basolateral part has the bidirectional link to the insular cortex and orbital cortex [18], [20], [21], [25], [26] and performs the main role in mediating memory consolidation [32] and providing the primary response, and is divided into three parts: the lateral, basal, and accessory basal [25], [29]. The lateral is the part through which stimuli enter the amygdala. The lateral region not only passes the stimuli to other regions, but also memorizes them to form the stimulus–response association [31]. This part also takes some roles in spreading the sensor's information to other parts, forming the association between the conditioned and unconditioned stimuli, inhabiting and reflecting the external stimuli, and memorizing the emotional experiences. The basal and accessory basal parts participate in mediating the contextual conditioning [20], [25].

The centeromedial part, which is the main output for the basaloteral part [26], is divided into the central and medial parts [18], [20], [25], [26]. It is responsible for the hormonal aspects of emotional reactions [25] or for mediating the expression of the emotional responses [25], [26].

*d) Orbitofrontal cortex* is located close to the amygdala and has a bidirectional connection to the amygdala. This part plays roles in processing stimulus [25], decoding the primary reinforcement, representing the negative reinforcement, and learning the stimulus–reinforcement association. It also evaluates and corrects reward and punishment [18]-[21], [33]-[37], selects goals, makes decisions for a quick response to punishment [18], [23], [25]-[36], and prevents inappropriate responses of the amygdala. The orbitofrontal cortex encompasses two parts, the medial and lateral. The medial part forms and memorizes reinforcement–stimulus association, and also has role in providing responses and monitoring them, whereas the lateral part evaluates the response and provides punishment [37].

*2) Emotion-Inspired Computational Models*

Computational models of emotional learning, which are computer-based models, have been developed to represent the associative learning aspect of emotional processing. From the application perspective, they can be categorized into three groups: emotion-based decision-making model, emotion-based controller, and emotion-based machine-learning approaches.

*a) Emotion-based decision-making model*: This model is the basis of artificial intelligent (AI) emotional agent that integrates emotional reactions with rational reactions. EMAI (Emotionally Motivated Artificial Intelligence) was one of the first attempts to develop emotion-based agents. It was applied for simulating artificial soccer playing [38], and its results were fairly good. The Cathexis model [39] was another emotional agent developed that reacted to an environment by imitating an emotional decision-making process in humans. The model of the mind [40] was developed as a modular artificial agent to generate emotional behavior for making decisions. An agent architecture that was called Emotion-based Robotic Agent Development (in reverse order, DARE) was developed on the basis of the somatic marker theory; it was tested in a multi-agent system and showed ability in modeling social and emotional behavior [41].

*b) Emotion-based controller*: The first practical implementation of an emotion-based controller is BELBIC (Brain Emotional Learning-Based Intelligent Controller) [7]. It was developed on the basis of Moren and Balkenius computational model [7], [23], [42]. The BELBIC has been successfully employed for a number of applications: controlling heating and air conditioning [43] of aerospace launch vehicles [44], intelligent washing machines [45], and trajectory tracking of stepper motor [46]. Another emotion-based intelligent controller is a neuro-fuzzy controller [47], which was integrated with emotion-based performance measurement to tune the parameters of the controller. Application of an emotion-based controller robotics was proposed in [48], which is an interesting example of applying emotional concepts in robotic applications and imitated the reinforcement learning aspect of emotional processing. The results of applying emotion-based controllers have shown that they have the capability to overcome uncertainty and complexity issues of control applications. Specifically, the BELBIC has been proven to outperform others in terms of simplicity, reliability, and stability [7], [39]-[42].

*c) Emotion-based machine-learning approach*: Developing machine-learning approaches by imitating emotional processing of the brain has captured the attention of researchers in the AI area. So far, some studies have been carried out to develop new neural networks by imitating some aspects of emotional learning. Hippocampus-neocortex and amygdala hippocampus model have been proposed as neural network models [49], [50]. They combine associative neural network with emotional learning concepts. Several emotion-based prediction models [6], [8]-[15] have been developed to model the complex systems. Most of them are based on the amygdala-orbitofrontal subsystem [23] that was proposed by Moren and Balkenius. They have been applied for different applications, e.g., auroral electrojec (AE) index prediction [9], solar activity prediction [6], [8], [10], [11]-[13], [14], and financial and chaotic time series prediction [6], [8]-[14].

*d) Amygdala-orbitofrontal system*

The fundamental model of emotion-based machine-learning approaches and emotion-based controllers is the amygdala-orbitofrontal system. It is a type of computational model of emotional learning with a simple structure that has been defined using the theory of the limbic system [23]. Its structure is inherited from some parts of the limbic system (e.g., the amygdala, thalamus, sensory cortex), and imitates the interaction between those parts of the limbic systems and formulates the emotional response using mathematical equations [23]. The amygdala-orbitofrontal subsystem consists of two subsystems: the amygdala and orbitofrontal subsystem. Each subsystem has several linear neurons and receives a feedback (a reward). The model's output function has been defined as subtracting the orbitofrontal's response from the amygdala's response. To update the weights, learning rules are defined for both the amygdala and orbitofrontal subsystem. Due to its simplicity, it has been the basis of most controllers and prediction models inspired by emotional learning.

### B. A Brief Overview of the Data- Driven Methods

One straightforward way to evaluate the data-driven learning approaches (e.g., neural network and neuro-fuzzy models) is to apply them to predict chaotic time series. Different types of neural network and neuro fuzzy models (e.g., RBF, ANFIS, LLNF) have been applied to model and predict the short-term and long-term behavior of chaotic time series (e.g., Lorenz, Henon, Mackey-Glass, Ikeda) [1]-[15], [51]-[55].

We have suggested a model that differs from the previously proposed models in terms of prediction accuracy, structural simplicity, and generalization capability. In the following, we explain the differences between BELPM and other well-known data-driven models.

*1) Radial Bias Function (RBF)* differs from BELPM in terms of the underlying structure, inputs of the neurons, connection between neurons, and number of learning parameters and learning algorithms.

*2) Generalization Regression Neural Network (GRNN)* [1] differs from BELPM in its number of neurons (i.e., the number of neurons of GRNN are equal to the size of training samples). Moreover, GRNN has no learning algorithm to optimize its performance and increase its generalization capability.

*3) Adaptive Neuro Fuzzy Inference System (ANFIS)* and BELPM are not similar because of different structures, functions, and some aspects of learning algorithms. Due to the learning algorithm and the large number of learning parameters (linear and nonlinear) that are spread through the layers, ANFIS has the capability to obtain very accurate results for complex applications. However, its learning algorithm has a significant effect on its computational complexity and it also causes over-fitting problems. The curse of dimensionality is another issue of ANFIS and increases the computational time of ANFIS for high-dimension application. Although the number of learning parameters of BELPM is not dependent on the dimension of input data, as mentioned before, BELPM uses Wk-NN; consequently, the computational time of BELPM only depends on the number of neighbors. To decrease its time complexity in high-dimension cases, we can choose the small number of neighbors for the BELPM.

4) *Local Linear Neuro Fuzzy Models (LLNF)* and BELPM can both be considered as types of "local modeling" [2] algorithms. They both combine an optimization-learning algorithm and LSE to train the learning parameters. However, LLNF uses Local Linear Model Tree (LoLiMoT) algorithm, instead of Wk-NN method of BELPM. The number of learning parameters of LoLiMoT has a linear relationship with the dimension of input samples and number of epochs; thus, its computational complexity has no exponential growth for high-dimension applications.

5) *Modular neural network* is a combination of several modules with different inputs [2] without any connection with others. There is no algorithm to update the learning parameters of the modules.

6) *Hybrid structures* that are defined in [1], differ from BELPM in receiving the input data. The submodules of a hybrid structure can also be designed in parallel or series.

III. BRAIN EMOTIONAL LEARNING-BASED PREDICTION MODEL

The architecture of BELPM is inspired by the brain emotional learning system. It mimics some functional and structural aspects of the limbic system regions. To describe BELPM's architecture and its underlying learning algorithms, we used the machine-learning terminology, instead of neuro-scientific terms. For example, an input–output pair of training data and an input–output pair of test data are equivalent to an unconditioned stimulus–response pair and a conditioned stimulus–response pair in neuro-scientific terminology, respectively. Thus, we used two subscripts $u$ and $c$ to distinguish the training data set and the test data set that are defined as $I_c = \{\mathbf{i}_{c,j}, r_{c,j}\}_{j=1}^{N_c}$ and $I_u = \{\mathbf{i}_{u,j}, r_{u,j}\}_{j=1}^{N_u}$, respectively, with $N_c$ and $N_u$ data samples. Before explaining the architecture of BELPM, let us briefly review the W-kNN algorithm that is the basic of BELPM. The following steps explain how the output value of an input vector $\mathbf{i}_{test}$ is calculated by using the W-kNN algorithm [56]:

1) Calculate the Euclidean distance as, $d_j = \left\| \mathbf{i}_{test} - \mathbf{i}_{u,j} \right\|_2$ for each $\mathbf{i}_{u,j}$ that is a member of the training data set $\{\mathbf{i}_{u,1}, \mathbf{i}_{u,2}, ..., \mathbf{i}_{u,N_u}\}$.

2) Determine the $k$ minimum values of $\mathbf{d} = \{d_1, d_2, ..., d_{N_u}\}$ as $\mathbf{d}_{min}$. The data samples corresponding to $\mathbf{d}_{min}$ are shown as $\mathbf{I}_{min} = \{\mathbf{i}_{min,j}, r_{min,j}\}_{j=1}^{k}$, where $\mathbf{I}_{min}$ denotes the samples of the training data set that are nearest neighbors to the test sample, $\mathbf{i}_{test}$.

3) *Calculate the output of $\mathbf{i}_{test}$ as given in (1).*

$$r_{test} = (\sum_{j=1}^{k} w_j \times r_{min,j} / \sum_{j=1}^{k} w_j) \quad (1)$$

The weight, $w_j$, is calculated as $w_j = K(d_j)$, where $K(.)$ is known as the kernel function that makes the transition from Euclidean distances to the weights. Any arbitrary function that holds the following properties given can be considered as the kernel function [56],[57]:

1) For all $d$, $K(d) \geq 0$.

2) If $d = 0$, then $K(d)$ gets the maximum value.

3) If $d \rightarrow \pm\infty$, then $K(d)$ gets the minimum value.

Some typical kernel functions are the Gaussian kernel (2), Inversion kernel (3), and the weighted kernel function that is defined as (4).

$$K(d) = \frac{1}{\sqrt{2\Pi}} exp(\frac{-d^2}{2}) \quad (2)$$

$$K(d) = \frac{1}{|d|} \quad (3)$$

$$K(d) = \frac{max(\mathbf{d}) - (d_j - min(\mathbf{d}))}{max(\mathbf{d})} \quad (4)$$

As mentioned earlier, they transform the Euclidean distances into the weights; thus, the neighbors that are closer to the test sample $\mathbf{i}_{test}$ have higher weights on estimating the output, $r_{test}$ [56],[57].

A. *Architecture of BELPM*

As shown in Fig. 2, the BELPM's architecture consists of four main parts that are named as TH, CX, AMYG, and ORBI, and are referred to as THalamus, sensory CorteX, AMYGdala, and ORBItofrontal, respectively.

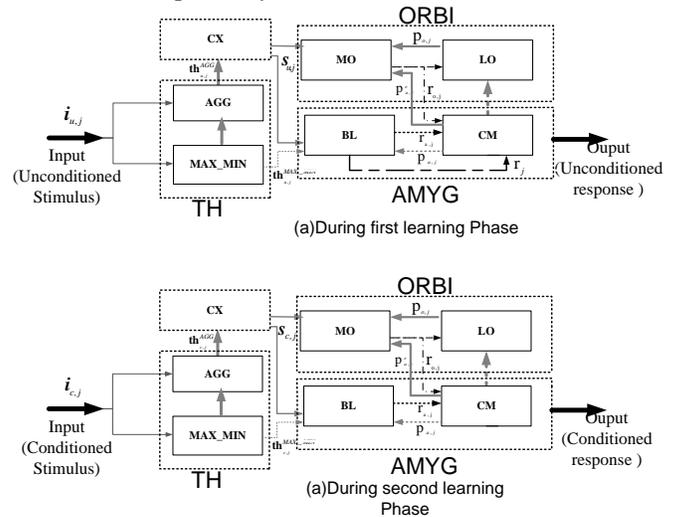

Fig.2. The architecture of BELPM showing the structure of each part and its connection to other parts. (a) An input from training set, unconditioned stimulus, enters the BELPM. (b) An input of test data, conditioned stimulus, enters the BELPM.

Let us assume an unseen input $\mathbf{i}_{c,j} \in I_c$ enters BELPM that provides the corresponding output using the following steps:

1) The input vector $\mathbf{i}_{c,j}$ is fed to TH, which is the entrance part of the BELPM structure. This part consists of two components: MAX_MIN (MAXimum_MINimum) and AGG

(AGGregation). The MAX_MIN can be described as a modular neural network. It has two neural networks, each of which has two layers with a competitive function for the neuron of the first layer and a linear function for the neuron of the second layer. The output of MAX_MIN that is referred to as $th_{c,j}^{MAX\_MIN}$ is calculated according to (5) and is fed to AGG and AMYG. Equation (5) calculates the highest and lowest values of the input vector with $R$ dimensions. The AGG can be described as a neural network with $R+2$ linear neurons ($R$ is the dimension of $i_{c,j}$); the output of AGG, $th_{c,j}^{AGG}$, is equal to $i_{c,j}$ and is fed to CX as shown in (6).

$$th_{c,j}^{MAX\_MIN} = [Max(i_{c,j}), Min(i_{c,j})] \quad (5)$$

$$th_{c,j}^{AGG} = [i_{c,j}] \quad (6)$$

2) The $th_{c,j}^{AGG}$ is sent to CX, which is a pre-trained neural network with one layer of linear function. The role of CX is to provide $s_{c,j}$ and distribute it between AMYG and ORBI. It should be noted that $i_{c,j}$ and $s_{c,j}$ have the same entity; however, they have been originated from different parts.

3) Both $s_{c,j}$ and $th_{c,j}^{MAX\_MIN}$ are sent to AMYG that is the main part of the BELPM structure and is divided into two components: BL(BasoLateral) and CM (CenteroMedial). The subpart that is referred to as BL corresponds to the set of the lateral and basal, while the other subpart, CM, corresponds to the accessory basal and centromedial part. There is a bidirectional connection between AMYG and ORBI; this connection is utilized to exchange the information, and contains AMYG's expected punishment and ORBI's response. The functionalities of AMYG have been defined to mimic some roles of the amygdala (e.g., storing unconditioned stimulus–response pairs, making the association between the conditioned and unconditioned stimuli, and generating reward and punishment). Thus, AMYG has a main role in providing the primary and final responses. The structure and function of BL and CM are given as follows:

*a) BL,* which is responsible for the provision of the primary response of the AMYG, calculates $d_{a,i} = \|s_{u,j} \cdot s_{c,j}\|_2 + \|th_{u,j} \cdot th_{c,j}\|_2$. Here, $s_{u,j}$ and $th_{u,j}^{MAX\_MIN}$ are the output of TH and CX for each member of the training data set $\{i_{u,1}, i_{u,2},...,i_{u,N_u}\}$, respectively. It must be noted that $s_{c,j}$ and $s_{u,j}$ are the output of the CX for $i_{c,j}$ and $i_{u,j}$, respectively. The BL encompasses an adaptive network with four layers (see Fig. 3(a)). The first layer consists of $k_a$ nodes ("adaptive or square" [3] nodes) with $K(.)$ function (kernel function). Each node has an input that is an entity from the $d_{amin} = [d_{amin,1}, d_{amin,2},...,d_{amin,k_a}]$ (which is a set of $k_a$ minimum distances of $d_a = [d_{a,1}, d_{a,2},...,d_{a,N_u}]$). The output vector of the first layer is $n_a^1$; the output value of each node of this layer is calculated using (7), where the input to $m^{th}$ node is $d_{amin,m}$.

$$n_{a,m}^1 = K(d_{amin,m}) \quad (7)$$

In general, the kernel function for $m^{th}$ node can be defined as (8), (9), and (10). The input and the parameter of $K(.)$ are determined using $d_m$ and $b_m$, which are the $m^{th}$ entity of $d$ and $b$, respectively. We used the subscript $a$ to distinguish BL's kernel function and its related parameters ($d_{amin}$ and $b_a$).

$$K(d_m) = exp(-d_m b_m) \quad (8)$$

$$K(d_m) = \frac{1}{(1+(d_m b_m)^{2z})} \quad (9)$$

$$K(d_m) = \frac{max(d) - (d_m - min(d))}{max(d)} \quad (10)$$

The second layer is a normalized layer and has $k_a$ nodes (fixed or circle), which are labeled as N to calculate the normalized value of $n_a^1$ as (11).

$$n_{a,m}^2 = \frac{(n_{a,m}^1)}{\sum_{m=1}^{k_a} n_{a,m}^1} \quad (11)$$

The third layer has $k_a$ circle nodes with functions given in (12). This layer has two input vectors, $n_a^1$ and $r_{ua}$; the latter is a vector that is extracted from $r_u = [r_{u,1}, r_{u,2},...,r_{u,N_u}]$ and is related to the $k_a$ nearest neighbors.

$$n_{a,m}^3 = \frac{(n_{a,m}^1)}{\sum_{m=1}^{k_a} n_{a,m}^1} \times r_{ua,m} \quad (12)$$

The fourth layer has a single node (circle) that calculates the summation of its input vector to produce $r_{a,j}$, the primary output (response). The function of the third and fourth layers can be formulated according to (13), i.e., the inner product of the third layer's output, $n_a^3$, and $r_{ua}$.

$$r_{a,j} = n_a^3 \bullet r_{ua} \quad (13)$$

The provided primary response, $r_{a,j}$, is sent to the CM to calculate the reinforcement signal. It should be noted that the connection between CM and BL (see Fig. 2(a)) does not exist between centromedial nuclei and the other parts of the amygdala. However, we assumed its existence in the BELPM's architecture.

*b) CM* is responsible to provide the final output; thus, it is an important component of AMYG. It has inputs from BL and ORBI, and performs different functions during the first leaning phase and the second learning phase of the BELPM.

- *CM during the first learning phase:* The first learning phase of BELPM begins when the input of BELPM is

chosen from the training sets, $i_{u,j} \in I_u = i_{u,1}, i_{u,2}, ..., i_{u,N_u}$ (see Fig. 2(a)). After receiving this input, BL starts performing its function and provides $r_{a,j}$ and $r_{u,j}$ that are sent to CM, which has three square nodes (see Fig. 3(b)). The function of the first node is defined according to (14) to provide $r_j$, i.e., the output of BELPM (emotional response). The functions of the second and third nodes are defined according to (15) and (16), which provide reinforcement (punishment), $p_{a,j}$, and expected reinforcement (expected punishment), $p_{a,j}^e$, respectively.

$$r_j = w_1 r_{a,j} + w_2 r_{o,j} + w_3 \tag{14}$$
$$p_{a,j} = w_{a,1} r_{u,j} + w_{a,2} r_{a,j} + w_{a,3} \tag{15}$$
$$p_{a,j}^e = r_{u,j} - r_{a,j} \tag{16}$$

A supervised learning algorithm determines appropriate values for the weights.

- *CM during the second learning phase*: The second learning phase of BELPM begins when the input is chosen from the test data sets $i_{c,j}$ (see Fig. 2(b)). As BL has no information about the desired output (the target response), it does not send any information about the desired output to the CM. In this phase, the first node of the CM has the same input and performs the same function as it does in the first learning phase. However, the input and the connection of the second square node are different. In this phase, the reinforcement, $p_{a,j}$, is calculated according to (17), which differs from (15) in terms of its input. The third node has no input and function, and it can be removed.

$$p_{a,j} = w_{a,1} r_j + w_{a,2} r_{a,j} + w_{a,3} \tag{17}$$

4) The expected reinforcements, $p_a^e$, and $s_{c,j}$ are sent to ORBI that is connected to CX and AMYG. The function of this part is defined to emulate certain roles of the orbitofrontal cortex. These roles include forming a stimulus–reinforcement association, evaluating reinforcement, and providing an output. Before explaining how ORBI performs, it should be noted that ORBI starts performing its functions after receiving the vector of the expected reinforcement, $p_a^e$, which means that BL of AMYG must have fulfilled its functions. The ORBI is composed of MO and LO corresponding to the lateral and medial parts of the orbitofrontal cortex, and their functions are described as follows:

a) *MO*, which is responsible for the provision of the secondary response of the BELPM, receives $s_{c,j}$ and calculates $d_{o,j} = \left\| s_{u,j} - s_{c,j} \right\|_2$ for each $s_{u,j}$. The MO consists of a four-layer adaptive network. The first layer has $k_o$ nodes (square), and the input vector and output vector of the first layer's nodes are $d_{omin}$ (the $k_o$ minimum values of the distance vector, $d_o = [d_{o,1}, d_{o,2}, ..., d_{o,N_u}]$) and $n_o^1$, respectively. The function of $m^{th}$ node is the kernel function given in (18).

$$n_{o,m}^1 = K(d_{omin,m}) \tag{18}$$

The second layer consists of $k_o$ nodes. Each node calculates an output as shown in (19) and sends it to the third layer.

$$n_{o,m}^2 = \frac{(n_{o,m}^1)}{\sum_{m=1}^{k_o} n_{o,m}^1} \tag{19}$$

The third layer has an input vector $p_{amin}^e$, which is a vector of $k_o$ minimum values of $p_a^e$ corresponding to $d_{omin}$. The third layer's nodes have the function to multiply $p_{amin}^e$ and $n_o^2$, as given in (20).

$$n_o^3 = n_o^2 \times p_{amin}^e \tag{20}$$

The fourth layer has a single node with a summation function that provides the output of ORBI (the secondary response). The result of the third and fourth layers can be obtained according to (21). The output $r_{o,j}$ is fed to the LO and CM.

$$r_{o,j} = n_o^2 \bullet p_{amin}^e \tag{21}$$

b) *LO* evaluates the output of MO, generates $p_{o,j}$ as reinforcement (punishment), and sends it to MO. It has one node (square) with a summation function given in (22).

$$p_{o,j} = w_{o,1} r_{o,j} + w_{o,2} \tag{22}$$

The main structure of BELPM is similar to the amygdala-orbitofrontal model. However, the connection between the components, their functions, and definition of reinforcement functions are significantly different from the amygdala-orbitofrontal model. Moreover, in this study, we have used pre-trained neural networks to explain the functionality of TH and CX; however, to increase the functionality and adaptability of BELPM, they can be defined by a multi-layer NN with trainable weights.

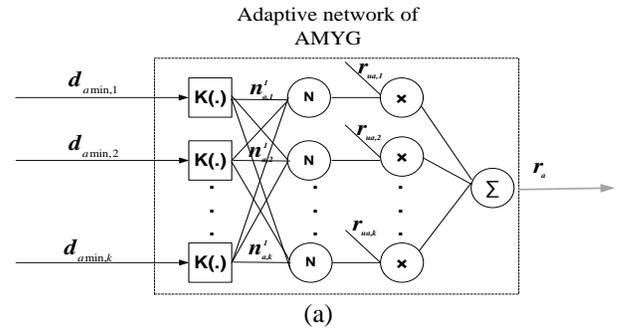

(a)

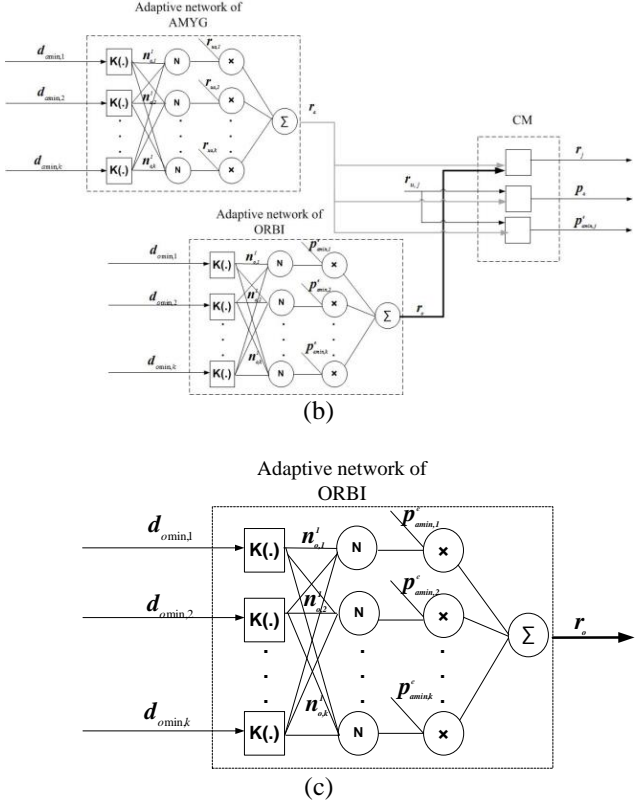

Fig. 3. (a). The adaptive network of BL. (b). The adaptive networks of AMYG and ORB and CM. (c). The adaptive network of ORBI.

## B. Learning Algorithms of BELPM

In the following, we explain how the BELPM uses the combination of two learning methods: the SD [3] and LSE to learn the input–output mapping (the stimulus–response association). The learning parameters are the weights ($w_1, w_2, w_3$, etc.) and the parameters of kernel functions ($b_o$ and $b_a$). As mentioned earlier, the learning algorithm of BELPM is divided into two phases: the first learning phase and the second learning phase. Each of them uses different learning rules to adjust the learning parameters.

*1) First learning phase:* At the first learning phase, a hybrid learning algorithm [3] that is a combination of SD and LSE is used to update the learning parameters of AMYG (e.g., $b_a$, $w_{a,1}, w_{a,2}$ and $w_1, w_2, w_3$) and ORBI ($b_o, w_{o,1}, w_{o,2}$). Under the assumption that the linear parameters, $w_1, w_2, w_3, w_{o,1}, w_{o,2}, w_{a,1}, w_{a,2}, w_{a,3}$ have fixed values, the nonlinear parameters (kernel parameters), $b_o$ and $b_a$, are updated by SD in a batch-learning manner. Thus, SD is applied to minimize the two loss functions, which have been defined on the basis of $p_a$ and $p_o$. Equations (23) and (24) are SD-based learning rules used to calculate the derivatives of the loss functions, with respect to $b_o$ and $b_a$.

$$b_a^{it+1} = b_a^{it} - \eta_a^{it} \times \nabla b_a^{it} \tag{23}$$

$$b_o^{it+1} = b_o^{it} - \eta_o^{it} \times \nabla b_o^{it} \tag{24}$$

The parameter $it$ denotes the current values of learning parameters, where $\nabla b_a^{it}$ and $\nabla b_o^{it}$ are the gradients of loss functions to the parameters $b_o$ and $b_a$ (25) and (26). Two learning rates $\eta_a^{it}$ and $\eta_o^{it}$ are defined as functions of $p_a$ and $p_a$.

$$\nabla b_a = \frac{\partial f(p_a)}{\partial b_a} \tag{25}$$

$$\nabla b_a = \frac{\partial f(p_a)}{\partial b_a} \tag{26}$$

where $p_a$ and $p_o$ are defined using (27) and (28).

$$p_a = w_{a,1} r_{u,a} + w_{a,2} r_a + w_{a,3} \tag{27}$$

$$p_o = w_{o,1} r_o + w_{o,2} \tag{28}$$

The ORBI and AMYG have their own loss function and update their own learning parameters $b_o$ and $b_a$ separately. The derivatives of (25) and (26) are formulated by using the chain rules of (29) and (30). Here, we have ignored $it$ and just mentioned about the chain rules, but in the algorithm, it has been calculated for the current iteration, $it$.

$$\frac{\partial p_a}{\partial b_a} = w_{a,2} \times r_a \times \frac{r_{u,a}}{(\sum_{m=1}^{k_a} n_{a,m}^1)^2} \times \frac{\partial n_a^1}{\partial K} \times d_{a\min} \tag{29}$$

$$\frac{\partial p_o}{\partial b_o} = w_{o,2} \times r_o \times \frac{r_{u,o}}{(\sum_{m=1}^{k_o} n_{o,m}^1)^2} \times \frac{\partial n_o^1}{\partial K} \times d_{o\min} \tag{30}$$

An offline version of LSE is used to update the linear parameters under the assumption that the nonlinear parameters have been updated and their values are fixed. The output of BELPM, $r_j$, is parameterized by the weights, $\{w_1, w_2, w_3\}$. Furthermore, $p_{a,j}$ and $p_{o,j}$ have been formulated using the linear parameters, $w_{a,1}, w_{a,2}, w_{a,3}, w_{o,1}$, and $w_{o,2}$, according to (27) and (28). The LSE updates the weights $w_1, w_2, w_3$, $w_{a,1}, w_{a,2}, w_{a,3}, w_{o,1}$, and $w_{o,2}$ by assuming that each triple of the set $\{(r_{a,j}, r_{o,j}, r_{u,j}), j=1,...N_u\}$ is substituted into (14).

Each triple of $\{(r_{a,j}, r_{u,j}, p_{a,j}), j=1,...N_u\}$ and each pair of $(r_{o,j}, p_{o,j}), i=1,...N_u$ are also substituted into (15) and (22), respectively; thus, $N_u$ linear equations such as (31), (32), and (33) are derived.

$$r_{u,j} = w_1 r_{a,j} + w_2 r_{o,j} + w_3 \tag{31}$$

$$p_{a,j} = r_{u,i} w_{a,1} + r_{a,i} w_{a,2} + w_{a,3} \tag{32}$$

$$p_{o,j} = r_{o,j} w_{o,1} + w_{o,2} \tag{33}$$

Equations (31), (32), and (33) can be rewritten in matrix form as $A = \left[ r_{a,j} r_{o,j} 1 \right]_{j=1}^{N_u}$, $B = \left[ r_{u,j} r_{a,j} 1 \right]_{j=1}^{N_u}$ and $C = \left[ r_{o,j} 1 \right]_{j=1}^{N_u}$ to define (34), (35), and (36), and update the linear parameters using LSE. Here, the weights are defined as $\mathbf{w} = [w_1, w_2, w_3]$, $\mathbf{w_a} = [w_{a,1}, w_{a,2}, w_{a,3}]$, and $\mathbf{w_o} = [w_{o,1}, w_{o,2}]$.

$$w = (A^T A)^{-1} A^T r_u \quad (34)$$
$$w_a = (B^T B)^{-1} B^T p_a^e \quad (35)$$
$$w_o = (C^T C)^{-1} C^T p_o^e \quad (36)$$

During the first learning phase, the learning parameters, linear and nonlinear, can be updated by using one of the following methods:
- All parameters can be updated using SD.
- The nonlinear parameters can be updated using SD and the initial values of linear parameters can be adjusted using LSE.
- The linear parameters are updated using LSE and the initial values of parameters of kernel functions are chosen by using a heuristic method.
- The nonlinear parameters are updated using SD and LSE are applied to update the linear parameters.

Certainly, these methods differ in terms of time complexity and prediction accuracy, and a tradeoff between high accuracy and low computational time must be considered to choose a feasible method. The batch mode or online mode of each method can be considered for the first learning phase.

*2) Second learning phase:* At the second learning phase, the nonlinear and kernel parameters are updated by a reinforcement-learning algorithm. The nonlinear parameters are updated by SD. The SD algorithm minimizes the loss functions, which are defined based on the reinforcements, $p_{a,j}$ and $p_{o,j}$, using (37) and (28). It must be noted that $p_a$ is calculated using the obtained output of BELPM, $r$, as given in (37), which differs from that calculated using (27). The adjusting rules that update $b_o$ and $b_a$ are calculated according to (21) and (22). The derivatives of the loss functions with respect to $b_o$ and $b_a$ are calculated according to (30) and (38). Here, the weights $w_{a,4}, w_{a,5}, w_{a,6}$ are equal to [1,-1,0].

$$p_a = w_{a,4} r + w_{a,5} r_a + w_{a,6} \quad (37)$$

$$\frac{\partial p_a}{\partial b_a} = w_{a,4} \times \frac{\partial r}{\partial b_a} + w_{a,5} \times r_a \times \frac{r_{u,a}}{(\sum_{m=1}^{k} n_{a,m})^2} \times \frac{\partial n_a^1}{\partial K} \times d_{amin} \quad (38)$$

## IV. CASE STUDIES: CHAOTIC TIME SERIES

In this section, BELPM is evaluated as a prediction model by using two benchmark chaotic time series, Lorenz and Henon. To provide a careful comparison with other methods, we used various data sets with different initialized points and sizes of training samples. We also utilized two error measures: normalized mean square error (NMSE) and mean square error (MSE), as given in (39), (40), to assess the performance of the prediction models and provide results comparable with other studies.

$$NMSE = \frac{\sum_{j=1}^{N}(y_j - \hat{y}_j)^2}{\sum_{j=1}^{N}(y_j - \bar{y}_j)^2} \quad (39)$$

$$MSE = \frac{1}{N} \sum_{j=1}^{N}(y_j - \hat{y}_j)^2 \quad (40)$$

Where $\hat{y}$ and $y$ refer to the observed values and desired targets, respectively. The parameter $\bar{y}$ is the average of the desired targets. For all experiments, one-fold cross-validation was chosen; the number of samples in one-fold cross-validation was equal to the size of the test data set.

*A. Lorenz Time Series*

The Lorenz time series [5], [9], [58] was chosen as the first test, given by using (41) and (42). In this case study, the initialized point is according to $x(0) = -15, y(0) = 0, z(0) = 0$.

$$\begin{aligned}\dot{x} &= a(y - x) \\ \dot{y} &= bx - y - xz \\ \dot{z} &= xy - cz \end{aligned} \quad (41)$$

$$a = 10, b = 28, c = 8/3, T = 0.01s \quad (42)$$

To produce the time series, the sampling period is equal to 0.01 s [5], [9], [58], and the embedded dimension is selected as three. The BELPM is tested for four sets of data from the Lorenz time series. The first data set is selected from 32[nd] to 51[st] s, and is employed for long-term prediction. For the training set, 500 samples are used and the next 1400 samples are considered as the test data set and validation data set. Table I presents the NMSEs obtained from applying different methods for 10, 30, and 40 steps ahead prediction of this data set. It also indicates that the NMSE of ANFIS is lower than the NMSE of BELPM for 10 steps ahead prediction. However, when the prediction horizon increases to 30 and 40 steps ahead, the NMSEs of BELPM are lower than ANFIS. The presented results in Table I also show that an increase in the prediction horizon causes a decrease in the prediction accuracy for all methods. It is important to note that the NMSEs of BELPM for predicting 30 and 40 steps ahead are less than those of the other methods.

For the second data set, ANFIS, Wk-NN, and BELPM are tested for 1–20 steps ahead prediction, using 1500 samples as the training data set and 1000 samples as the test data set. In Fig. 4, the NMSE values vs. the prediction steps are depicted; it can be observed that the NMSE values of BELPM are lower than those of the other methods, especially, in the case of 15 steps ahead prediction and longer.

TABLE I
THE COMPARISONS OF NMSEs OF DIFFERENT METHODS TO PREDICT MULTI-STEP AHEAD OF LOREN TIME SERIES (THE FIRST DATA SET)

| Learning Method | Ten | Thirty | Forty |
|---|---|---|---|
| BELPM | 0.0125 | 0.2473 | 0.2447 |
| ANFIS | 0.006 | 0.3559 | 0.3593 |
| RBF | 0.4867 [58] | 0.3405 | 0.6887 |
| LLNF | 0.1682 [58] | 0.4946 | 0.5341 |
| Wk-NN | 0.0235 | 0.2599 | 0.3830 |

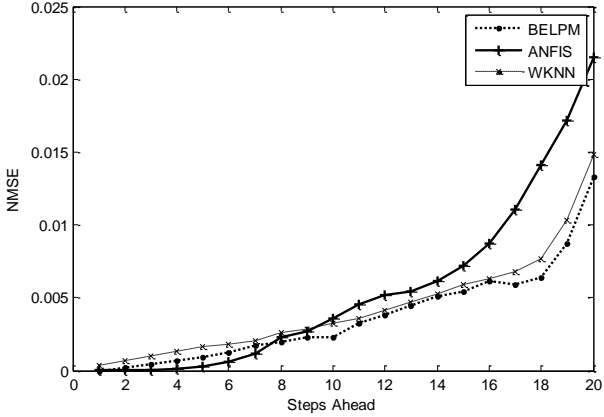

Fig.4. The NMSE values of multi-step ahead prediction of Lorenz system vs. prediction horizon for 1000 samples of test data using 1500 samples as training data (the second data set). The dotted curves show the NMSE of BELPM.

In the third data set, ANFIS, Wk-NN, and BELPM are compared for 30 steps ahead prediction, with different sizes of training data set. In this case, the complete sample set is chosen from $30^{th}$ to $55^{th}$ s, and the training is conducted using 500, 1000, and 1500 samples to predict 1000 samples of test data set from $45^{th}$ to $55^{th}$ s. The validation set is 1000 samples from $56^{th}$ to $65^{th}$ s. Table II compares the values of NMSE for the different methods and once again shows that NMSE of BELPM has the lowest value among that of all the methods tested.

For the fourth data set, the training set and test data are picked from $15^{th}$ to $45^{th}$ s and $46^{th}$ to $55^{th}$ s, respectively. BELPM is applied for 25 steps ahead prediction and the NMSE values, CPU time, and structures of different models are compared in Table III. Again, it can be observed that the NMSE of BELPM is lower than that of the other methods. The time complexity of BELPM is less than most of the other methods, and only Wk-NN is found to have a little lower time complexity than BELPM.

TABLE II
THE COMPARISONS OF NMSEs OF DIFFERENT METHODS FOR THIRTY STEPS AHEAD PREDICTION OF LORENZ TIME SERIES USING DIFFERENT SIZES OF TRAINING SAMPLES.

| Learning Method | 500/1000 | 1000/1000 | 1500/1000 |
|---|---|---|---|
| BELPM | 0.1711 | 0.0108 | 0.0280 |
| ANFIS | 0.4199 | 0.1002 | 0.0947 |
| Wk-NN | 0.1951 | 0.1518 | 0.0425 |

TABLE III
THE COMPARISONS OF NMSEs, CPU TIME AND STRUCTURE OF DIFFERENT METHODS TO PREDICT 25 STEPS AHEAD USING 3000 TRAINING DATA SAMPLES

| Learning Method | NMSE | TIME(Sec) | STRUCTURE |
|---|---|---|---|
| BELPM | 0.0325 | 11.81 | 10 neuron |
| ANFIS | 0.0802 | 12.91 | 4 rule |
| LoLiMoT | 0.2059 | 18.74 | 7 neuron |
| RBF | 0.1193 | 26.07 | 32 neuron |
| Wk-NN | 0.0342 | 1.23 | 5 neighbor |

As mentioned earlier, the Lorenz chaotic time series is a well-known benchmark time series and has been tested with numerous data-driven models to evaluate the models' performance. Table IV presents the obtained NMSEs of several data-driven methods for noiseless and noisy data. The data-driven models are: Nonlinear Autoregressive model with eXogenous input (Hybrid NARX-Elman RNN) [59], Evolving Recurrent Neural Networks (ERNN) [60], Radial Basis Function (RBF), multilayer perceptron (MLP) [5], Support Vector Regression (SVR), Tapped Delay Line Multilayer Perceptron (TDL-MLP), Distributed Local Experts based on Vector_Quantization using Information Theoretic learning (DLE-VQIT) [61], Cooperative Coevolution of Elman Recurrent Neural Networks (CCRNN) [62], Functional Weights Wavelet Neural Network-based state-dependent AutoRegressive (FWWNN-AR) [63], Recurrent Neural Network trained with Real-time Recurrent Learning (RNN-RTRL), Recurrent Neural Network trained with the second-order Extended Kalman Filter (RNN-EKF), Recurrent Neural Network trained with the algorithm and BackPropagation Through Time (BPTT), feedforward Multi layer Perceptron trained with the Bayesian Levenberg–Marquardt (MLP-BLM), and recursive second-order training of Recurrent Neural Networks via a Recursive Bayesian Levenberg–Marquardt (RBLM-RNN) algorithm [65]. It can be noted that the table is sorted according to the obtained NMSEs, and that the NMSE of BELPM is not excellent as that of the other methods. However, this table compares the NMSEs for short-term prediction, which is not a feasible application for BELPM.

*B. Henon Time Series*

The second benchmark of this study is the Henon time series that is constructed by using (43).

$$x(t+1) = 1 - ax(t)^2 + y(t)$$
$$y(t+1) = bx(t) \qquad (43)$$
$$a = 1.4, b = 0.3$$

The embedded dimension are considered as 0.01s [5], [58] and three, respectively. In this case study, the initialized point is $x(0)=0, y(0)=0$. Three data sets of the Henon time series are used for evaluating the BELPM. The first data set is selected from $9^{th}$ to $18^{th}$ s and the training data set and the test data set consist of 800 and 100 samples, respectively. The BELPM, ANFIS, and Wk-NN are tested to predict three steps ahead of this data set.

TABLE IV
THE COMPARISONS OF NMSEs OF DIFFERENT METHODS FOR LORENZ TIME SERIES

| Learning Method | NMSE | No of Training and Test | Fundamental Method and steps ahead |
|---|---|---|---|

|  |  | data samples |  |
|---|---|---|---|
| FWWNN[63] | 9.8e-15 | 1500,1000 | NN+AR, 1 step noiseless |
| NARX[59] | 1.9e-10 | 1500,1000 | AR, 1 step noiseless |
| BELRFS[12] | 4.9e-10 | 1500,1000 | NF+BEL, 1 step noiseless |
| ERNN[60] | 9.9e-10 | 1500,1000 | NN, 1 step noiseless |
| RBF[5] | 1.4e-9 | 1500,1000 | NN, 1 step noiseless |
| MLP[5] | 5.2e-8 | 1500,1000 | NN, 1 step noiseless |
| **BELPM** | 2.9e-6 | 1500,1000 | W-kNN+BEL,1step noiseless |
| TDL-MLP[61] | 1.6e-4 | ----- | NN, 1 step noiseless |
| DLE-VQIT[61] | 2.6e-4 | ----- | -----, 1 step noiseless |
| LSSVMs[64] | 6.4e-5 | 1000,250 | NF, 1 step noisy-STD 0.05 |
| **BELPM** | 3.7e-4 | 1500,1000 | W-kNN+BEL,1stepnoisy-STD 0.001 |
| **BELPM** | 4.4e-4 | 1500,1000 | W-kNN+BEL,1stepnoisy-STD 0.01 |
| RBLM-RNN[65] | 9.0e-4 | 1000,250 | RNN,1 step noisy-STD 0.05 |
| CCRNN [62] | 7.7e-4 | 500,500 | NN,2 step noiseless |
| LLNF[64] | 2.9e-4 | 1000,250 | RNN, 1 step noisy-STD 0.05 |
| MLP_BLM[65] | 8.1e-4 | 1000,250 | NN, 1 step noisy-STD 0.05 |
| MLP_EKF[65] | 1.6e-3 | 1000,250 | NN, 1 step noisy-STD 0.05 |
| RNN-RTRL[65] | 1.7e-3 | 1000,250 | NN, 1 step noisy-STD 0.05 |
| RNN-BPTT[65] | 1.8e-3 | 1000,250 | NN, 1 step noisy-STD 0.05 |
| RNN-EKF[65] | 1.2e-3 | 1000,250 | RNN,1 step noisy-STD 0.05 |
| **BELPM** | 1.0e-3 | 1500,1000 | W-kNN+BEL,1stepnoisy-STD 0.1 |

Table V presents the NMSEs, structure, and computational time (CPU time) for the tested methods. It can be noted that the NMSE of BELPM is lower than that of the other methods. This table also indicates that the number of neurons in BELPM is more than the number of rules in ANFIS; however, the CPU time for BELPM is lesser than that of ANFIS. The graph in Fig. 5 displays the obtained errors (the difference between the target values and obtained outputs) from ANFIS and BELPM for the test data set. We have also compared the effect of different structures of W-kNN, ANFIS, and BELPM on their prediction accuracy. The structures of these methods have been changed by increasing the number of neighbors in BELPM and Wk-NN and the number of rules in ANFIS. As Fig. 6 shows the value of NMSE for ANFIS decreases when the number of rules increases. In contrast, increasing the number of neighbors of Wk-NN increases the value of NMSE. For BELPM, the value of NMSE decreases slowly when the number of neighbors increases. This experiment verifies the generalization capability of BELPM and shows that different structures of BELPM do not make a noticeable difference between the obtained prediction errors. To further evaluate the performance of the BELPM and verify its robustness, white noise with standard deviation 0.1 is added to the first data set, and the results of multi-step ahead prediction are listed in Table V. The effect of using the second learning phase of BELPM has been presented in Table VI. It is clear that there is a reduction in the obtained NMSEs of BELPM because of the second learning phase (SLP).

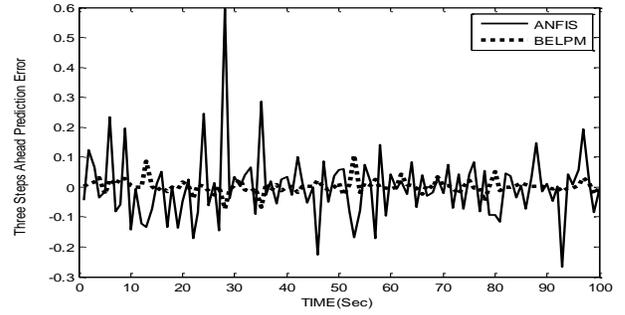

Fig. 5. The obtained error of three steps ahead predictions of the Henon time series by BELPM and ANFIS.

TABLE V
THE COMPARISONS OF NMSEs, CPU TIME AND STRUCTURE OF DIFFERENT METHODS TO PREDICT THREE STEPS AHEAD OF HENON TIME SERIES (FIRST DATA SET OF HENON)

| Learning Method | NMSE | Structure | CPU Time(Sec) |
|---|---|---|---|
| BELPM | 0.0065 | 27neuron | 0.3315 |
| RBF[5] | 0.0872 | 42 neuron | ------- |
| LoLiMoT[5] | 0.0291 | 20 neuron | ------- |
| ANFIS | 0.0232 | 25rule | 153.3 |
| WKNN | 0.0107 | 5 neighbor | 0.0725 |

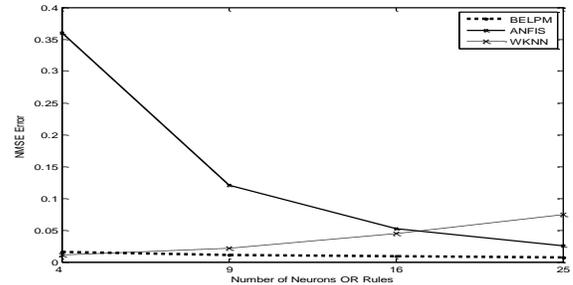

Fig. 6. The comparison of NMSEs of three different structures obtained using the BELPM, ANFIS, and W-kNN. These methods have been used for different structures.

TABLE VI
THE COMPARISONS OF NMSEs OF DIFFERENT METHODS TO PREDICT MULTI-STEPS AHEAD OF HENON TIME SERIES

| Learning Method | 1Step-ahead | 2Step-ahead | 3Step-ahead |
|---|---|---|---|
| BELPM(SLP) | 0.0063 | 0.0251 | 0.1436 |
| BELPM(FLP) | 0.0066 | 0.0274 | 0.1576 |
| ANFIS | 0.0051 | 0.0244 | 0.2539 |
| LoLiMoT | 0.0590[5] | 0.2475 | 0.7048 |
| WKNN | 0.0078 | 0.0479 | 0.1897 |

The graph in Fig. 7 shows the values of NMSE of two steps ahead prediction of the first data set during the learning phases. It can be observed that the NMSE decreases continuously during the first and second learning phases. The graph in Fig. 8 depicts the mean square error (MSE) that is obtained from the training samples and test samples during the first learning phase. In each epoch, the learning parameters have been updated and the values of MSE for the training data and the test data have been calculated. It is noticeable that the values of the dotted curve that is related to the MSE of the test data are lower than those of the solid curve, i.e., the MSE of training data.

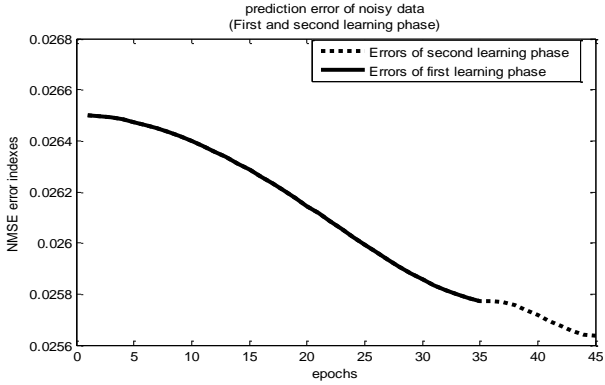

Fig.7. The NMSE of BELPM to predict the two steps ahead of Henon time that is added white noise with standard deviation 0.1. The solid line is related to the NMSE during 35 epochs of the first learning phase and the dashed line is the NMSE during 10 epochs of the second learning phase.

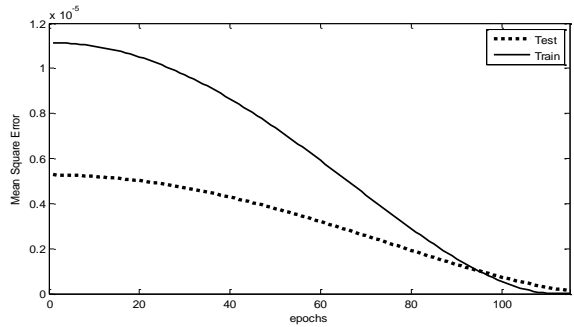

Fig.8. Curves of the mean square error during the first learning phase; the solid line is the MSE curve, which is related to training data; the dashed line is the MSE curve, which is related to test data.

The second data set of the Henon time series is selected from $8^{th}$ to $18^{th}$ s, and three different sizes of data samples are selected as training data. Table VII compares the obtained results from examining different methods that are trained by different training sets: 300, 600, and 900 training samples, and have been used to predict 100 samples of test data. As shown in Table VII, the prediction accuracy of BELPM is higher than that of the other methods in all cases; thus, it is a suitable method for the prediction application with a small number of training samples. The bar chart in Fig. 9 shows the NMSE for the compared methods. It can be seen that the values of NMSE of BELPM are once again lowest among those of all the presented methods. Examination of this data set revealed that a decrease in the number of training samples causes an increase in the values of NMSE of all methods (e.g., BELPM, ANFIS, Wk-NN, and RBF). However, the rate of increasing the values of NMSE of BELPM is lower than that of other methods. The third data set of the Henon time series is selected to compare the time complexities (CPU time) of various learning methods when the size of the training set is large. A training set of 3000 samples are selected from $100^{th}$ to $130^{th}$ s, and the following 1000 samples from $130^{th}$ to $140^{th}$ s are considered as the test data set. Table VIII summarizes the NMSEs, CPU time, and the applied structures of the different methods. Although the CPU time of Wk-NN is 1.128 s, which is less than that of BELPM, the prediction accuracy of BELPM is higher than that of Wk-NN and other methods.

TABLE VII
THE COMPARISONS OF NMSEs OF DIFFERENT METHODS TO PREDICT 100 DATA SAMPLES USING DIFFERENT NUMBER OF NUMBER OF TRAINING SAMPLES

| Method | 900/100 | 600/100 | 300/100 |
|---|---|---|---|
| BELPM | 0.0067 | 0.0102 | 0.0269 |
| ANFIS | 0.0301 | 0.0406 | 0.0538 |
| WKNN | 0.0107 | 0.0118 | 0.0468 |
| RBF | 0.0155 | 0.0307 | 0.0457 |

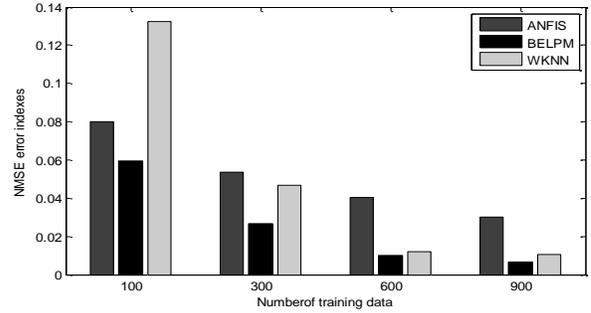

Fig. 9. The NMSE of different methods for three steps ahead prediction of Henon time series using different sizes of training data (100, 300, 600, and 900) for 100 samples of test data.

TABLE VIII
COMPARISON OF DIFFERENT METHODS TO PREDICT THREE STEP AHEAD OF HENON TIME SERIES FOR THIRD SET OF DATA

| Learning Method | NMSE | TIME(Sec) | STRUCTURE |
|---|---|---|---|
| BELPM | 0.0077 | 5.72 | 5neuron |
| ANFIS | 0.0514 | 627.56 | 16 rules |
| LLNF | 0.2059 | 18.74 | 7 neuron |
| RBF | 0.1193 | 26.07 | 32 neuron |
| WKNN | 0.0083 | 1.128 | 3 neighbor |

Figure 10 displays the obtained errors from examining BELPM and ANFIS for three steps ahead prediction. The figure shows that BELPM is more accurate than ANFIS. Figure 11 depicts how the NMSE index of BELPM decreases over its learning phases. It becomes obvious that for both training and test data, there is a continuous decrease in the NMSE index during the epochs. For this experiment, we have applied BELPM for the third data set and have chosen 200 epochs for both first and second learning phases. This figure verifies the ability of BELPM to adjust the learning parameters in the online mode.

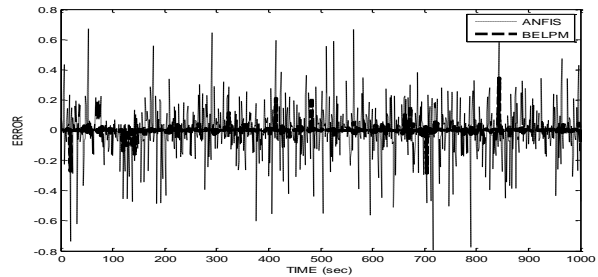

Fig.10. The prediction errors of ANFIS and BELPM for three steps ahead prediction of Henon time series using 3000 training samples.

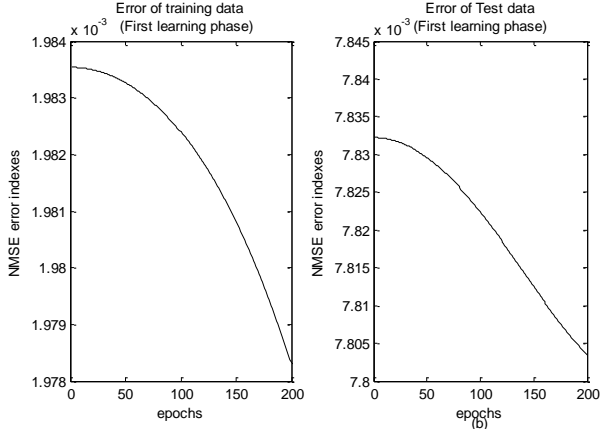

Fig.11. The NMSE of three steps ahead predictions during first learning phase. (a) For training data. (b) For test data.

The Henon chaotic time series is also a well-known benchmark time series, and has been examined by a large number of data-driven models. Table IX lists the obtained NMSEs of several data-driven methods. It becomes obvious that BELPM has a fairly good performance in predicting noisy data time series.

TABLE IX
THE NMSE VALUES OF DIFFERENT METHODS TO PREDICT HENON TIME SERIES

| Learning Method | NMSE | No of Training and Test data samples | Fundamental Method and time series |
|---|---|---|---|
| RBF[5] | 1.4e-9 | 1500,1000 | NN, 1 step noiseless |
| MLP[5] | 5.2e-8 | 1500,1000 | NN, 1 step noiseless |
| TDL-MLP[61] | 1.6e-4 | ----- | NN, 1 step noiseless |
| DLE-VQIT[61] | 2.6e-4 | ----- | -----, 1 step noiseless |
| **BELPM** | 3.3e-4 | 800,100 | W-kNN+BEL, 1 step noisy-STD 0.001 |
| LSSVMs[64] | 4.4e-4 | 1000,250 | NF, 1 step noisy-STD 0.05 |
| RBLM-RNN[65] | 6.8e-4 | 1000,250 | RNN,1 step noisy-STD 0.05 |
| RNN-EKF[65] | 8.6e-4 | 1000,250 | RNN,1 step noisy-STD 0.05 |
| LLNF | 7.7e-4 | 1000,250 | RNN, 1 step noisy-STD 0.05 |
| MLP-BLM[65] | 8.1e-4 | 1000,250 | NN, 1 step noisy-STD 0.05 |
| **BELPM** | 6.4e-3 | 800,100 | W-kNN+BEL ,1 step noisy-STD 0.01 |
| RNN-BPTT[65] | 1.1e-3 | 1000,250 | NN, 1 step noisy-STD 0.05 |
| RNN-RTRL[65] | 1.0e-3 | 1000,250 | NN, 1 step noisy-STD 0.05 |

## V. DISCUSSION AND CONCLUSION

This study has presented a prediction model inspired by brain emotional processing and, particularly, has investigated this model for chaotic time series prediction. We have described the architecture of this model using feedforward neural networks and adaptive networks. Furthermore, we have also explained the function and learning algorithms of the model that is referred to as BELPM. The accuracy of the BELPM has been extensively evaluated by different data sets of two benchmark chaotic time series, Lorenz and Henon, and the results strongly indicate that the model can predict the long-term state of chaotic time series rapidly and more accurately than other well-known methods, i.e., RBF, LoLiMoT, and ANFIS. The results also show that BELPM is more efficient than the other methods when large training data sets are not available.

In comparison with other data-driven approaches, we can summarize the highlighted features of BELPM as follows:

*1)* It converges very fast into its optimal structure (see Tables III, IV, and VII).

*2)* It has high generalization capability (see dotted lines in Figs. 7 and 8), high robustness from the perspective of learning algorithms (see solid lines in Figs. 7, 8, 11, and 12), low noise sensitivity (see Table V), relatively low computational time (compare the values of time columns in Tables III, IV, and VII), and low model complexity.

*3)* The number of neighbors in AMYG and ORBI is not dependent on the input's dimension, which indicates that the dimension of data does not have a direct effect on the model complexity. However, the input's dimension increases the computational time of calculating the Euclidean distance and searching the nearest neighbors; thus, it has an indirect effect on the computational time of BELPM. The experiments have also shown that when there are a limited number of samples for training the model, the number of neighbours, $k_o$ and $k_a$ should be raised to get accurate results (compare Table IV and VII). A general observation is that a better prediction result could be achieved when the number of neighbours, $k_o$, in the ORBI part is approximately twice the number of neighbours, $k_a$, in the AMYG part. The feasible values and effective combination of $k_a$ and $k_o$ can be determined using meta-heuristic optimization methods (e.g., genetic algorithm).

*4)* The second learning phase of BELPM provides online adaptation and continuously increases the prediction accuracy and offers the capability to overcome both over-fitting and under-fitting problems.

*5)* For long-term prediction using a small number of training samples, the accuracy of BELPM is higher than other data-driven models, such as ANFIS and LoLiMoT (see Tables I, II, IV, and V). In the case of using a large number of training samples, it is also noticeable that the computational time of BELPM is not greater than that of ANFIS and LoLiMoT (see Tables III and VII). It can also be concluded that when the degree of chaos is high, even using a large number of data and neuro-fuzzy methods (ANFIS and LoLiMoT) would not achieve higher prediction accuracy than the BELPM (see Table VII).

As future works, the authors consider adding some optimization methods (e.g., genetic algorithm) to find optimal values of the fiddle parameters, e.g., the number of neighbors $k_a$ and $k_o$ and the initial values of nonlinear parameters. Other improvements in the model would be made on the basis of kd-Tree data structure [47] to address "the curse of dimensionality" [1] problem and decrease the computational time complexity of BELPM. To adjust the nonlinear parameters, different types of optimization methods (e.g., Quasi-Newton or Conjugate Directions) for ORBI and AMYG can be utilized. In addition, the Temporal Difference (TD) learning algorithm can also be used as a reinforcement method for the second learning phase to update the linear learning parameters. The good results obtained by employing the BELPM for predicting the chaotic time series are a motivation for applying this model as a classification method as well as to identify complex systems.